\documentclass[%
 reprint,
 amsmath,amssymb,
aps,
prl,
]{revtex4-1}

\usepackage{graphicx}
\usepackage{dcolumn}
\usepackage{bm}
\usepackage{amssymb}

\begin{document}


\title{On the role of system size in Hall MHD magnetic reconnection}

\author{C. Bard}
\email{christopher.bard@nasa.gov}
\author{J. C. Dorelli}%
\email{john.dorelli@nasa.gov}
\affiliation{%
 NASA Goddard Space Flight Center	
}%

\date{\today}

\begin{abstract}
We study the effects of the Hall electric field on magnetic island coalescence in the large island limit and find evidence for both a elongated electron current sheet layer with a Sweet-Parker-like reconnection rate and a collapsed, Petschek-like electron sheet with a peak reconnection rate approaching the $0.1v_AB_0$ Hall MHD rate. The state observed in our simulations appears to depend on grid scale. Furthermore, even at the largest system sizes, we find that flux-pileup effects cause the islands to ``bounce'' despite the presence of a collapsed current sheet allowing for fast instantaneous reconnection. The average reconnection rate in the large island limit is slow though the peak reconnection rate is fast. 
\end{abstract}

\maketitle


Magnetic reconnection is an ubiquitous process, yet it remains poorly understood. Despite studying this mechanism for over sixty years, we are still trying to completely understand and explain how it works across a variety of length scales \cite{zweibel16}.
In this Letter, we use a Hall MHD code \cite{bard14,bard16ph} to undertake a systematic study of how Hall MHD reconnection in the island coalescence problem changes with island size and numerical resistivity.

Flux pileup reconnection is a well-studied problem in resistive MHD (e.g. \cite{biskamp80,knoll06}), and has recently been studied in Hall MHD \cite{dorelli03,dorelli03b,knoll06b,simakov09,zocco09,stanier15,stanier16} and kinetic codes \cite{karimabadi11,stanier16}.
The resistive MHD simulations demonstrate that flux pileup causes the reconnection electric field to become independent of resistivity until some maximum pileup limit is reached \cite{biskamp80,wang96}.
In this ``saturation'' limit, however, the reconnection rate becomes heavily dependent on resistivity and so-called ``sloshing'' motions, in which the islands continually bounce back and forth as they coalesce \cite{clark64,craig99,knoll06}.

Adding the Hall term to the resistive MHD simulations introduces another important length scale to the problem, the ion inertial length $d_i$.
\cite{dorelli03} demonstrated the existence of a ``whistler-mediated'' regime in which $d_i > \ell_\eta$, $\ell_\eta$ being the length scale associated with resistivity ($\eta$).
\cite{knoll06b} showed that although the Hall term allows the reconnection rate to become independent of resistivity, the pileup mechanism still induces a strong system size dependency.
On the other hand, \cite{stanier15} did not observe sloshing in their Hall MHD simulations and found a maximum reconnection rate independent of system size.

More recent particle-in-cell simulations \cite{karimabadi11,stanier16} show sloshing of large islands, though their maximum reconnection rates appear to be insensitive to system size.
Additionally, the largest islands in \cite{karimabadi11} bounced once and then stopped coalescing.

Here, we seek to understand these conflicting results by exploring the complex relationship between resistive scales ($\ell_\eta$), Hall scales ($d_i$), and global scales (here represented by the parameter $\lambda$).
Our simulations span a large range of grid scales and island sizes relative to $d_i$, though the combination of the smallest grid scales and largest island sizes are, at the moment, too computationally expensive.

Using an explicit finite-volume Hall MHD code based on the algorithm of \cite{toth08}, we simulate the Fadeev equilibrium \cite{fadeev65} in the $xy$-plane which describes an island chain with vector potential $A_z(x,y) = -\lambda B_0 \ln[\cosh(y/\lambda)+\epsilon\cos(x/\lambda)]$ and magnetic field $\mathbf{B} = \mathbf{\nabla}\times A_z \mathbf{\hat{z}}$, with $B_0$ the asymptotic magnetic field. 
For all simulations in this paper, we set $\epsilon = 0.4$ and define the simulation domain such that $L_x = 4\pi\lambda$ and $L_y = L_x/2$ with $x\in [-L_x/2, L_x/2]$ and $y\in[-L_y/2, L_y/2]$.
We control the island size through varying the parameter $\lambda$.
Boundary conditions are periodic in $x$ and conducting in $y$.

To maintain MHD equilibrium ($\nabla p = \vec{J}\times\vec{B}$), we define a pressure profile $p = 0.5(1-\epsilon^2)/[\cosh(y/\lambda)+\epsilon\cos(x/\lambda)]^2 + p_b$ and assume a constant ideal gas temperature such that the density is given by $\rho = p$. 
We set the background pressure $p_b = 0.5 p_0$.

Following \cite{daughton09,karimabadi11}, we disturb this equilibrium with in-plane magnetic perturbations $\tilde{B_x} = 0.5\delta_B (L_x/L_y)\cos[2\pi(x-L_x/2)/L_x]\sin[\pi y/L_y]$ and $\tilde{B_y} = -\delta_B\sin[2\pi(x-L_x/2)/L_x]\cos[\pi y/L_y]$ with $\delta_B = 0.1 B_0$.

We normalize density to a reference density $\rho_0$, the magnetic field to the asymptotic field strength $B_0$, and velocities to a reference Alfv\'en velocity, $v_{A} = B_0/\sqrt{4\pi \rho_0}$.
Pressure is normalized such that $p_0 = B_0^2/(4\pi)$.
In order to better compare the simulations across the range of island sizes, we normalize the time to the global Alfv\'en time scale $t_A = L_x/v_{A0}$.
We set the reference length to the ion inertial length, $L_0 = d_i$, such that the normalized $\bar{d}_i = d_i/L_0 = 1$.

Our Hall MHD code does not have a resistivity term, so we cannot enforce an uniform $\eta$ across the simulation domain.
Throughout this paper, we use the grid scale ($\Delta x,~\Delta y$) as a proxy for resistivity (c.f. \cite{toth08}, equation 35).

We simulated a grid of runs with scales $0.033\leq \Delta x, \Delta y\leq 0.2$ in the range $6.04 \leq \lambda \leq 25.46$; some island sizes were run with even finer grid meshes ($\Delta x \to 0.012$).
We selected a output snapshot nearest $0.84 t_A$ for all runs, which is close to the time of peak reconnection rate for each run.
The reconnection rate is measured as the out-of-plane electric field $E_z = E_R = \frac{\partial}{\partial t}[A_{yX} - A_{yO}]$ along the $y=0$ axis, with $A_{yX}$ ($A_{yO}$) the magnetic potential measured at the X-point (O-point).
$E_R$ is normalized in units of $v_{A0} B_0$.

\begin{figure}
\includegraphics[width=0.49\textwidth]{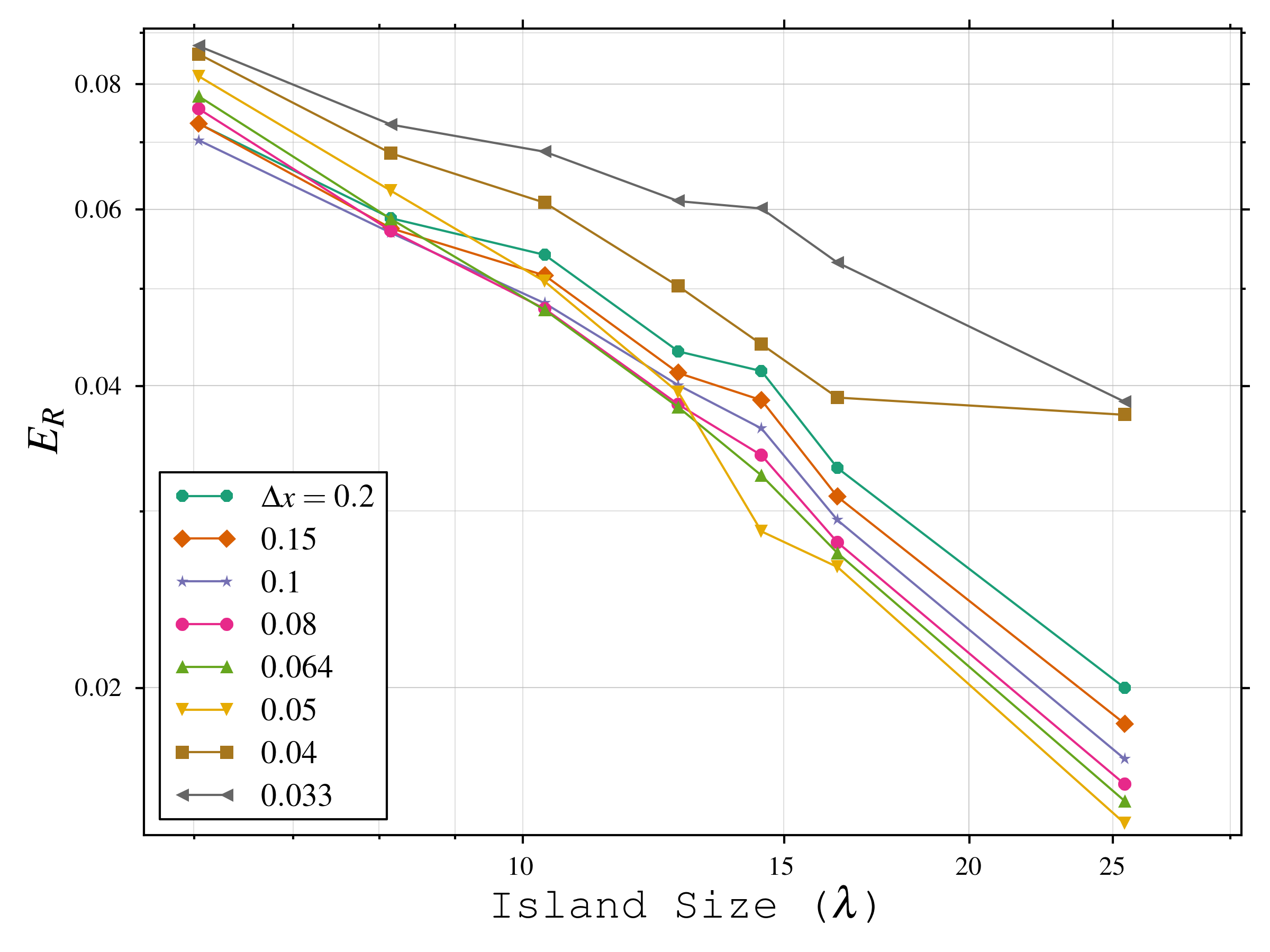}
\caption{\label{fig:er_lam}Peak Hall MHD reconnection rate ($E_R$) vs. island size ($\lambda$) for several different grid spacings ($\Delta x$). All units are normalized. Despite the influence of the Hall term, the reconnection rate still depends on the system size. This is likely a consequence of flux-pileup and sloshing.}
\end{figure}

\begin{figure}
\includegraphics[width=0.49\textwidth]{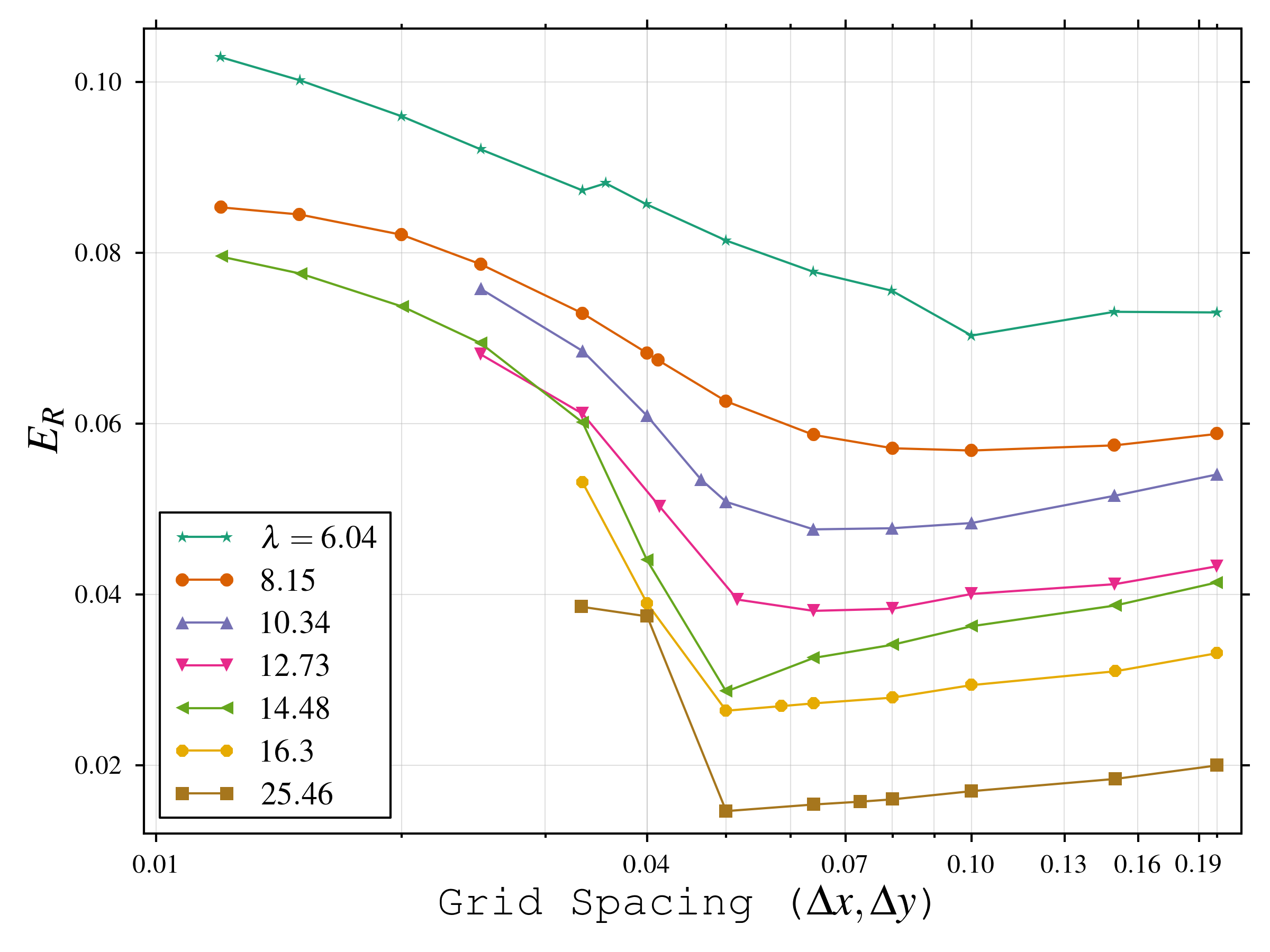}
\caption{\label{fig:er_dx}Peak Hall MHD reconnection rate ($E_R$) vs. grid spacing ($\Delta x$) for several different island sizes ($\lambda$). All units are normalized. The large jump in $E_R$ for $\Delta x \leq 0.04$ is coincident with the collapse of the current sheet (c.f. Figure \ref{fig:bx_current})}.
\end{figure}

First, we corroborate the results of \cite{dorelli03b,knoll06b}, who found evidence for slow reconnection in Hall MHD: including the Hall term does not guarantee that reconnection will be fast.
Figures \ref{fig:er_lam} and \ref{fig:er_dx} show the reconnection rate as a function of both system size and grid scale for each simulation.
These plots illustrate two key results: 1) at larger grid scales, Hall MHD reconnection exhibits a Sweet-Parker-like system-size dependency; 2) at sufficiently small grid scales, reconnection becomes faster than Sweet-Parker and may eventually approach the system-size-independent $0.1 v_A B_0$ rate found by \cite{shay99}.

For our simulations, the dividing line between these two regimes occurs within $0.04 \lesssim \Delta x \lesssim 0.05$.
This suggests that there is some critical level of resistivity which determines the structure of the current sheet and the resulting type of reconnection. 
At the moment, it is uncertain if this dividing line is independent of $\lambda$ or would exhibit a system-size dependency.

\begin{figure}
\includegraphics[width=0.49\textwidth]{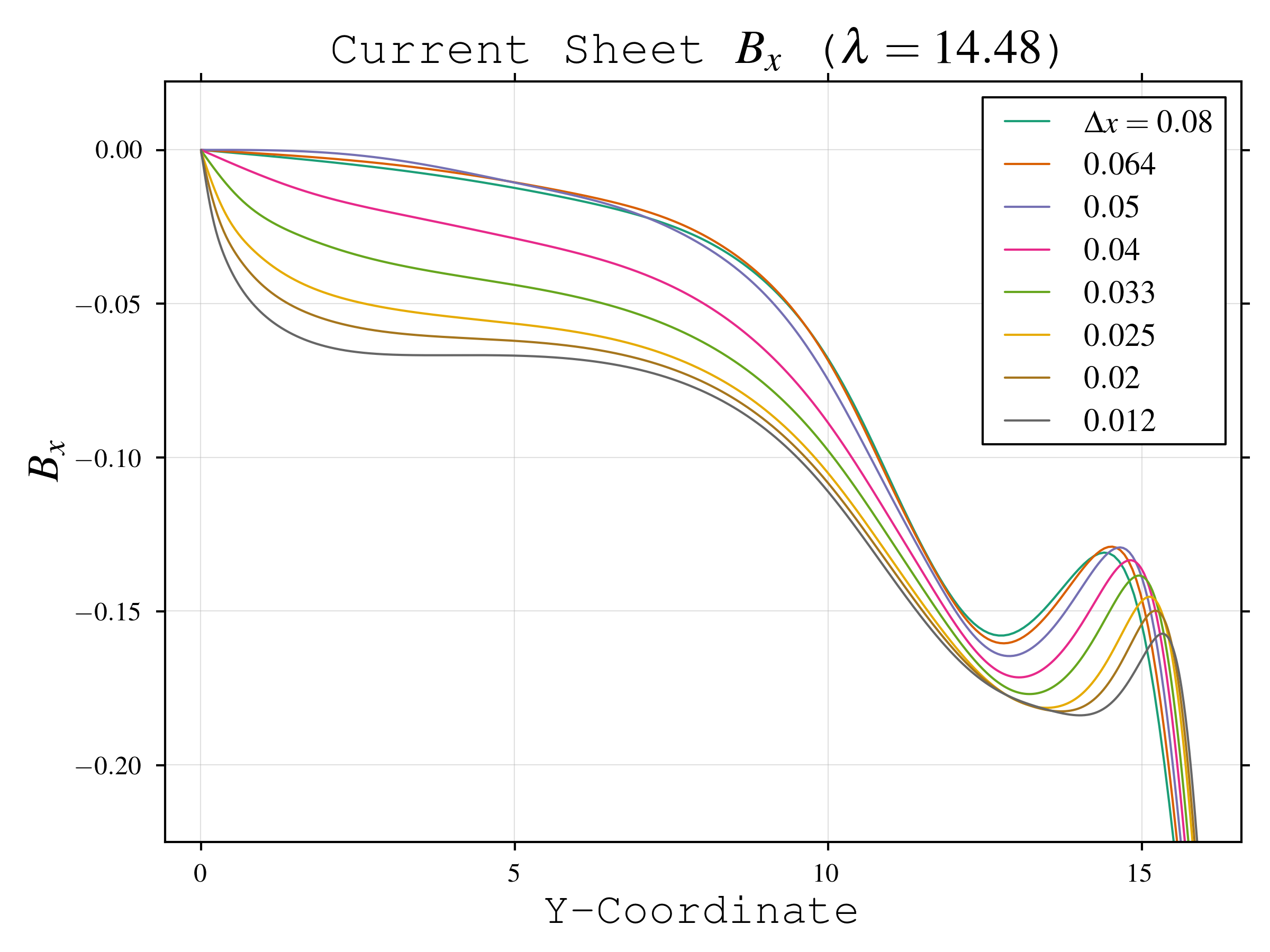}
\caption{\label{fig:bx_current} In-plane magnetic field $B_x$ along the center of the current sheet (measured along the y-axis) at $t\approx 0.85 t_A$ for island size $\lambda = 14.48$. At grid scales $\Delta x \leq 0.04$, the sheet structure changes from a single scale to two scales.}
\end{figure}

The abrupt jump in reconnection rate at $\Delta x \leq 0.04$ is coincident with a structural change in the current sheet (Figure \ref{fig:bx_current}). For $\Delta x > 0.04$, $B_x$ appears to be relatively ``flat'' before it starts to match the background field external to the islands.
However, with sufficiently low resistivity, the current sheet collapses to a two-scale structure.
The ``flat'' scale persists in the outer (further from the center) region, but there is now a sharp boost to $B_x$ in some small inner scale.
As the resistivity (or grid scale) decreases, the magnitude of this boost increases simultaneously with a shrinking inner scale length.
Since $B_x$ is stronger, the $\vec{J}\times\vec{B}$ acceleration boosts the outflow velocity along the length of the current sheet (in $\hat{y}$).
This, along with the decreasing inner scale length, allows for a faster reconnection rate.

\begin{figure}
\includegraphics[width=0.49\textwidth]{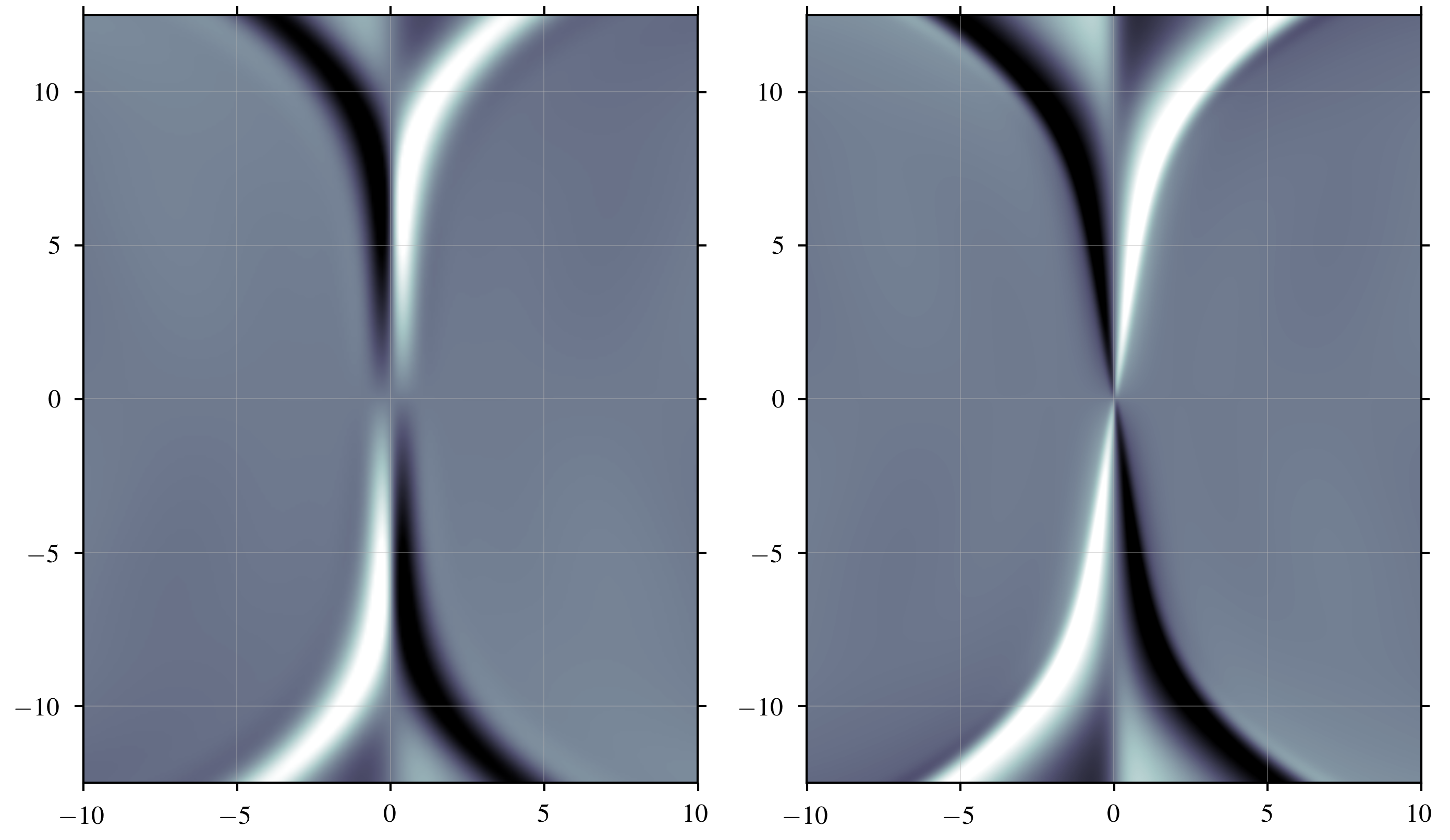}
\caption{\label{fig:bz_14} Zoom-in of out-of-plane magnetic field $B_z$ at $t \approx 0.85 t_A$ for island size $\lambda = 14.48$ with a grid scale $\Delta x = 0.08$ (left) and $\Delta x = 0.025$ (right). The current sheet structure of the larger grid scale is long and narrow, representative of a Sweet-Parker sheet. When the grid scale becomes sufficently small, the current sheet collapses and forms a Petschek-like configuration.}
\end{figure}

An additional structural change in the current sheet can be seen in the out-of-plane quadrupolar field. 
Figure \ref{fig:bz_14} illustrates $B_z$ at $t \approx 0.85 t_A$ with an island size $\lambda = 14.48$ for two different grid scales, $\Delta x = 0.08$ and $0.025$.
The larger grid scale (left side of Fig. \ref{fig:bz_14}) shows the uncollapsed sheet, while the smaller grid scale (right of same figure) demonstrates the collapsed sheet structure.
The uncollapsed sheet is more Sweet-Parker-like: a narrow and extended ion-inertial layer.
On the other hand, the collapsed sheet is shorter and wider, appearing to be a Petschek-like configuration with a larger outflow opening angle.
We observe that the opening angle increases as the grid scale decreases.

This finding may resolve some debate in the literature concerning the behavior of reconnection in the Hall MHD model.
\cite{shay99,shay01,bhattacharjee05}, with Hall MHD codes, demonstrated that Hall reconnection is fast in conjunction with a short electron dissipation region.
However, some kinetic models (e.g. \cite{daughton06}) found that the electron layer actually stretches along the outflow direction, which leads to a flow bottleneck which throttles the reconnection rate.
In response, \cite{shay07} presented particle-in-cell simulations showing a clear two-scale structure of the electron region (see their Figure 4) with fast reconnection, corroborating earlier results.

We find both behaviors in our simulations: the elongated, Sweet-Parker-like current sheet appears at larger $\Delta x$ and the two-scale, Petschek-like sheet appears at smaller $\Delta x$.
This may be explained by the ``bistability'' model of \cite{cassak10}, in which resistivity determines whether reconnection is slow, fast, or capable of both.
We postulate that a smaller grid scale in our simulations decreases the resistive scale length while increasing the maximum whistler wave speed.
The competition between these two physical processes could be what sets the current sheet behavior and resulting reconnection rate.
With a sufficiently small grid scale, the dominance of dispersive waves could allow the peak reconnection rate to approach the universal $E_R\sim 0.1~V_AB_0$ rate found by \cite{shay99} even at large system sizes, though ideal MHD sloshing effects will limit the duration of this peak and lower the overall average reconnection rate.

\begin{figure}
\includegraphics[width=0.49\textwidth]{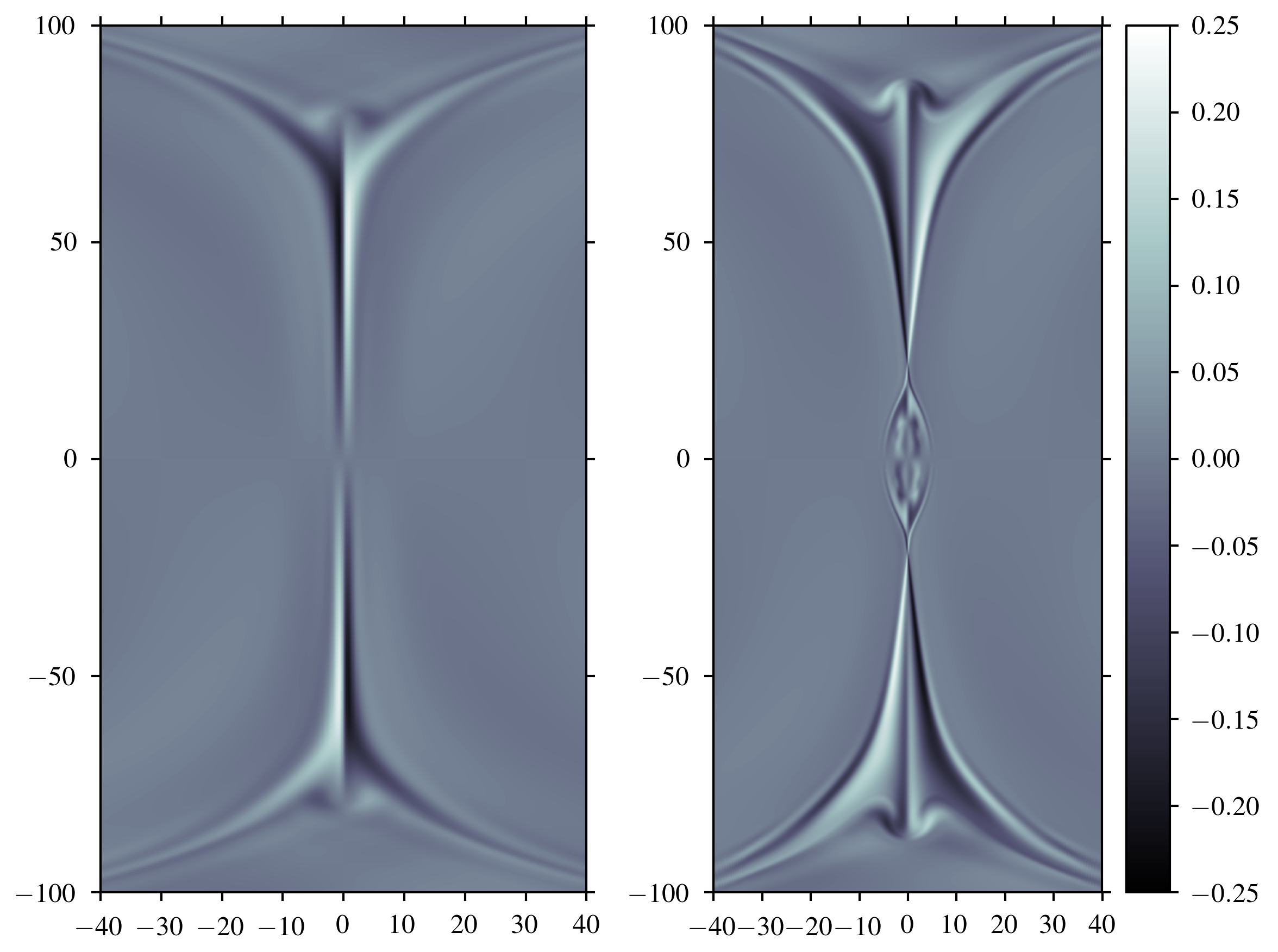}
\caption{\label{fig:lam95_sheet} Zoom-in of out-of-plane magnetic field $B_z$ at $t \approx 0.84 t_A$ for island size $\lambda = 95.37$ with grid scales $\Delta x = 0.1$ (left) and $\Delta x = 0.04$ (right). A Sweet-Parker sheet is seen on the left, and a Petschek-like sheet is shown on the right.}
\end{figure}

The effect of resistivity on reconnection persists even at the largest island size we simulated: a coalescence run with $\lambda = 95.37$ with $\Delta x = 0.04$ demonstrates a flared, Petschek-like current sheet, while the lower-resolution ($\Delta x = 0.1$) run has a Sweet-Parker-type sheet (Figure \ref{fig:lam95_sheet}).
Furthermore, both of the $\lambda = 95$ island simulations exhibited bouncing, like the kinetic simulations of \cite{karimabadi11}.

Bouncing appears to be caused by the competition between pileup and reconnection: the former piles up magnetic flux even as the latter works to clear it out.
Flux pileup is a ideal MHD effect; adding kinetic effects or the Hall term to this picture simply changes the physics of the current sheet and how quickly magnetic flux can reconnect.
In our Hall MHD simulations with $\Delta x = 0.1$, the onset of bouncing occurred for island sizes $\lambda \approx 10 d_i$ while similar ideal MHD simulations showed bouncing for $\lambda \gtrsim 3 d_i$. 
The Hall-boosted reconnection rate is able to clear out more flux, so islands have to be larger before pileup induces bouncing.

\begin{figure}
\includegraphics[width=0.49\textwidth]{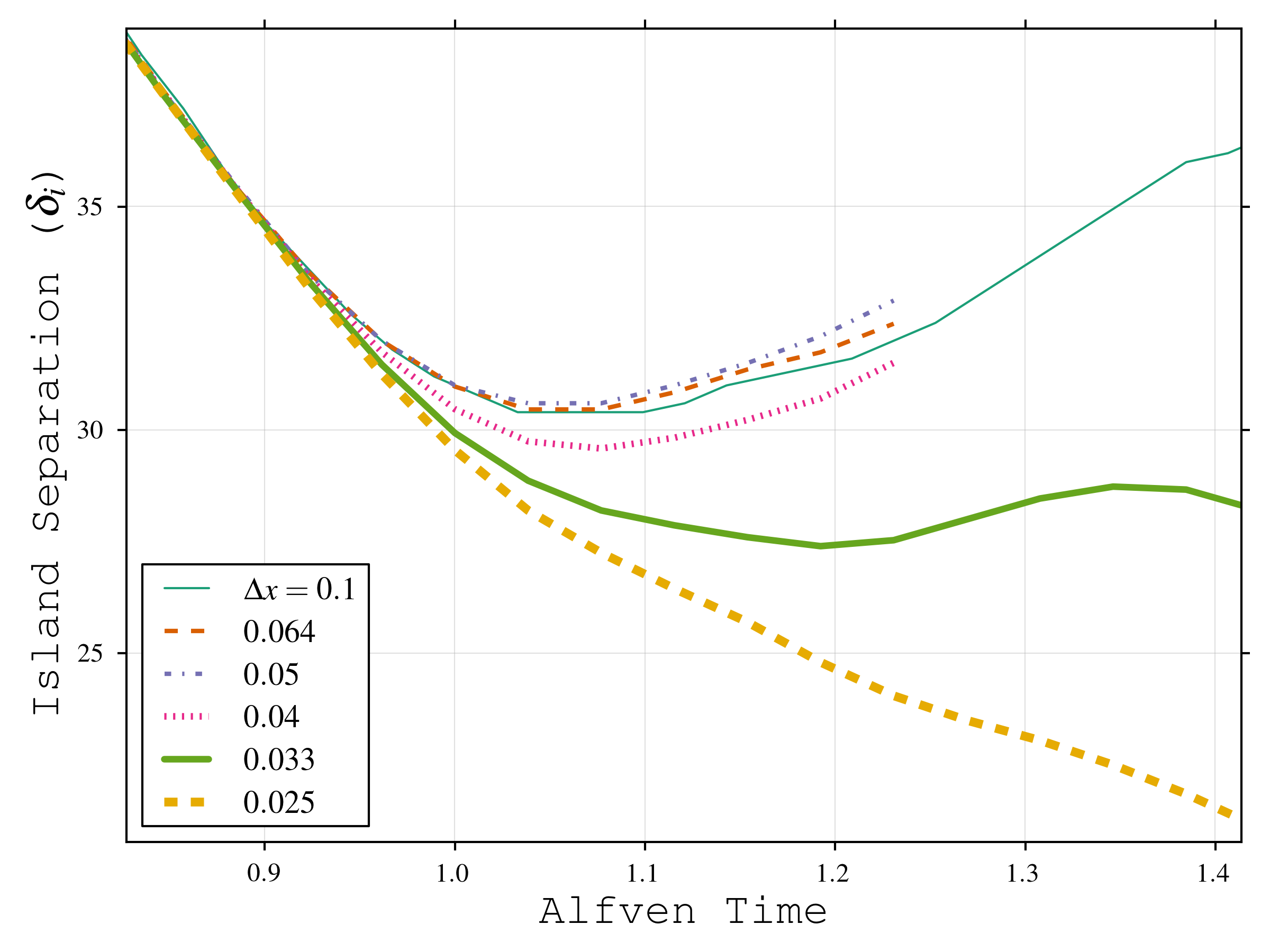}
\caption{\label{fig:Lsep} Island separation as measured between the center of the two islands over time; data is taken from selected $\lambda = 14.48$ runs. Although the current sheet changes structure at scales $\Delta x \lesssim 0.04$, bouncing persists until $\Delta x \lesssim 0.025$.}
\end{figure}

Resistivity also influences Hall MHD bouncing by controlling the reconnection rate.
Figure \ref{fig:Lsep} illustrates the gradual evolution of sloshing for $\lambda = 14.48$ at several different grid scales.
As the grid scale gets smaller and reconnection faster, sloshing weakens and the bouncing eventually disappears for $\Delta x \lesssim 0.025$.
Interestingly, bouncing can still occur even with collapsed current sheets ($\Delta x < 0.05$).
Since the reconnection rate appears to saturate at $0.1 V_A B_0$ \cite{shay99}, this implies that there may be an island size beyond which reconnection cannot prevent flux pileup and sloshing, even in the limit of zero resistivity.

\begin{figure}
\includegraphics[width=0.49\textwidth]{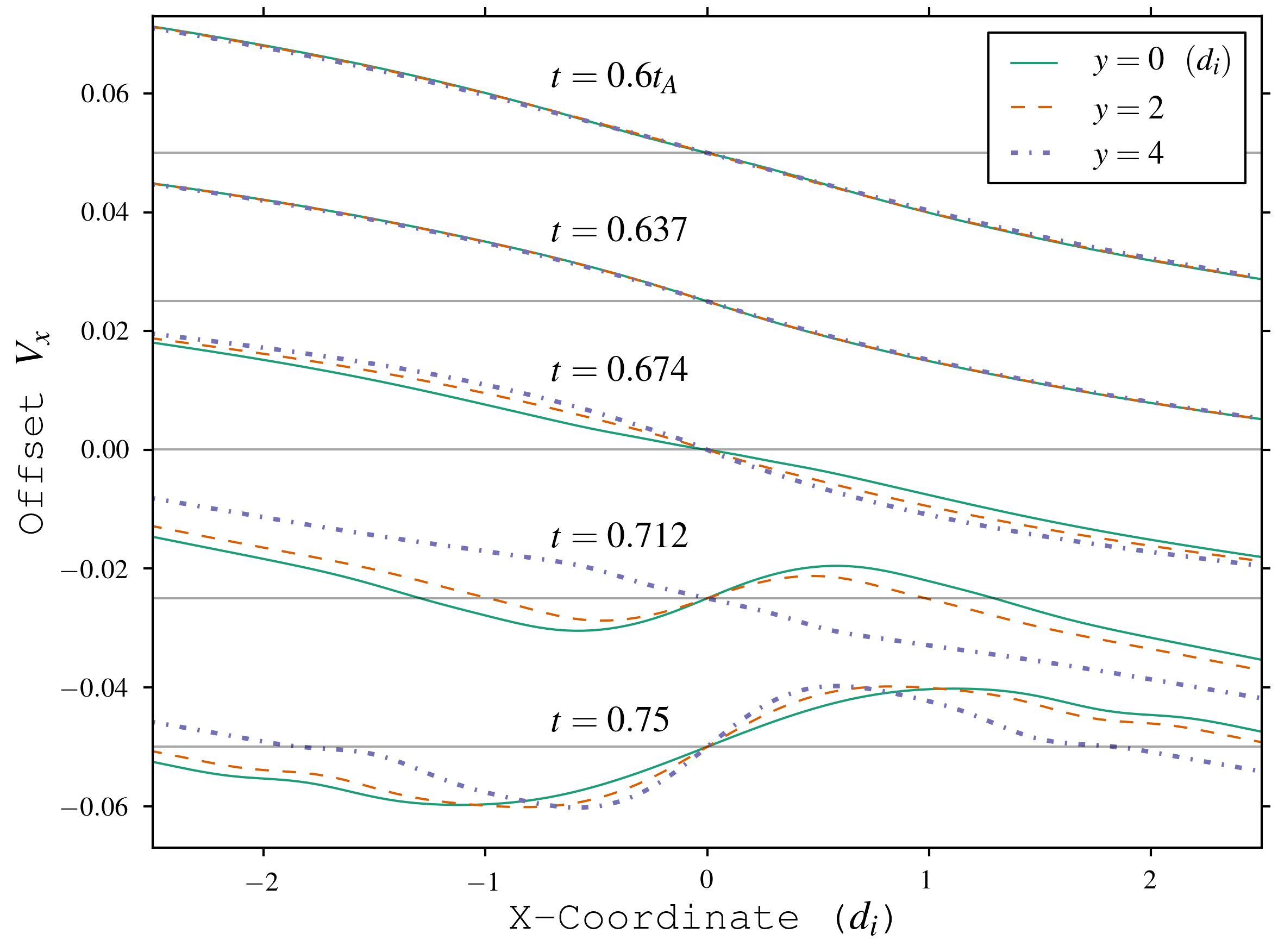}
\caption{\label{fig:vx_island} Velocity parallel to island motion ($\hat{x}$) for selected times prior to and after onset of bounce for $\lambda = 95.37$; the offset for each time is marked by light gray lines. The uneven pileup and resulting $\vec{J}\times \vec{B}$ force results in non-uniform deceleration along the $y$-axis where the islands meet.}
\end{figure}

We note that both $\lambda= 95$ current sheets ($\Delta x = 0.1$ and $0.04$) formed sub-islands; these were not present in the more intermediate-size island runs (e.g. Figure \ref{fig:bz_14}).
While the cause of these sub-islands is not yet certain, we postulate that, in general, they are caused by perturbations in the repulsive $\vec{J}\times \vec{B}$ force induced by uneven pileup across the current sheet.
This causes multiple reconnection sites to form between the islands, creating sub-islands when outflows from adjacent reconnection sites oppose one another.
For the first bounce, however, there is more pileup at the center of the islands ($y=0$) than further away, causing a larger deceleration in $v_x$ there (Figure \ref{fig:vx_island}).
We hypothesize that the central piled-up flux was sufficient to prevent reconnection at $y=0$, but not in the adjacent cells.
In the Sweet-Parker case ($\Delta x = 0.1$), the slower reconnection outflows formed a sub-island too small to stop the overall bouncing.
As a result, other sub-islands were able to form as the outer edges of the islands piled up prior to the full bounce.
Conversely, for $\Delta x =0.04$, a single large sub-island formed between two Petschek-like reconnection sites and the islands bounced apart more quickly.

As island size extends beyond 100 $d_i$, the onset of reconnection may migrate towards the edges of the coalescing islands as flux pileup increases at the center.
Eventually, in this simple picture, it may be possible for large enough islands to be ``too big to coalesce''.
This will require further study, though at such large scales the relatively thin current sheets will likely be unstable to the plasmoid instability \cite{loueriro07,cassak09,daughton09,huang13,comisso16} or an ideal tearing mode \cite{pucci14,landi15, pucci17}. 

Interestingly, sub-islands did not appear in large-island ($\lambda \sim 100$) PIC simulations \cite{karimabadi11}.
We speculate that ion gyro-viscosity, which is not in Hall MHD, could reduce  momentum shear caused by non-uniform bouncing; this could prevent adjacent reconnection sites from forming in such close proximity and interacting to form sub-islands.

In sum, our results suggest that when reconnection is externally driven by large magnetic structures, the dissipation physics (e.g., size of effective resistivity) plays a critical role in determining whether we get a rapid, bursty release of energy or a very low reconnection rate.
The presence of strong Hall electric fields does not ensure fast and instantaneous reconnection.

Our results suggest that the level of resistivity affects the structure of the current sheet and the resulting reconnection rates.
We speculate that this inconsistent Hall MHD behavior might be consistent with the Cassak et al. ``bistability'' model \cite{cassak10} and that there is some critical resistivity at which we suddenly transition from an extended ion inertial sheet (with $E_R\propto\lambda$) to a more Petshcek-like configuration (with $E_R$ weakly dependent on $\lambda$).

Finally, island sloshing could be a possible contributor to the storage of magnetic energy before the catastrophic onset of solar flare reconnection \cite{uzdensky07,cassak08,shepherd10,cassak12}. 
Although reconnection can be faster than Sweet-Parker, the flux pileup prevents a sustained, steady-state reconnection event in which magnetic energy is completely dispersed.
Eventually, as the overall system evolves, enough flux has dissipated such that pileup no longer prevents an explosive transition into a prolonged state of fast reconnection.

\section*{acknowledgements}
CB and JD acknowledge the use of the NASA NAS (pleiades) and NCCS (discover). CB acknowledges support from the NASA Postdoctoral Program at GSFC.

\bibliography{island}
\end{document}